\documentclass[a4paper,10pt,nofootinbib]{revtex4}%{article}
\usepackage[utf8]{inputenc}
\usepackage[english]{babel}
\usepackage{amsmath,amssymb}
\usepackage{comment}
\usepackage{graphicx}
\usepackage{url}
\usepackage{color}

\newcommand{\sigmatot}{\sigma_\text{tot}}
\newcommand{\sigmael}{\sigma_\text{el}}
\newcommand{\sigmadiff}{\sigma_\text{diff}}
\newcommand{\sigmasd}{\sigma_\text{SD}}
\newcommand{\sigmadd}{\sigma_\text{DD}}
\newcommand{\bbC}{\mathbb{C}}

\DeclareMathOperator{\Real}{\text{Re}}
\DeclareMathOperator{\Imag}{\text{Im}}
\begin{document}

\title{A model of diffractive excitation  in $pp$ collisions at high energies}
\author{M. Broilo}
\email{mateus.broilo@ufrgs.br}
\affiliation{High and Medium Energy Group, \\
Instituto de F\'{\i}sica e Matem\'atica, Universidade Federal de Pelotas\\
Caixa Postal 354, CEP 96010-900, Pelotas, RS, Brazil}
\author{V.~P. Gon\c{c}alves}
\email{barros@ufpel.edu.br}
\affiliation{High and Medium Energy Group, \\
Instituto de F\'{\i}sica e Matem\'atica, Universidade Federal de Pelotas\\
Caixa Postal 354, CEP 96010-900, Pelotas, RS, Brazil}
\author{P. V. R. G. Silva}
\email{pvrgsilva@ufpel.edu.br}
\affiliation{High and Medium Energy Group, \\
Instituto de F\'{\i}sica e Matem\'atica, Universidade Federal de Pelotas\\
Caixa Postal 354, CEP 96010-900, Pelotas, RS, Brazil}
\date{\today}
\date{\today}

\begin{abstract}
A phenomenological model for the description of the single and double diffractive excitation in $pp$ collisions at high energies is presented. Considering the Good -- Walker approach, we propose a model for the eigenstates of the scattering operator and for  the treatment of the interaction between them, with the high energy behavior of the cross section driven by perturbative QCD. The behavior of the total, elastic, single and double diffractive cross sections are analyzed and predictions for the energies of Run 3 of the LHC and those of the Cosmic Rays experiments are derived. We demonstrate that the model describes the current data for the energy dependence of the cross sections. A comparison with the recent data for the $\rho$ parameter and the differential elastic cross section are also presented and shortcomings of the current model are discussed.

\end{abstract}

\maketitle

\section{Introduction}
\label{sec:intro}
One of the more basic problems in the theory of strong interactions -- the Quantum Chromodynamics (QCD) -- is the description of the total, inelastic, elastic and diffractive hadronic cross sections. At the same time, it is  the more challenging one, due to the dominance of the nonperturbative sector of the theory and the difficulties associated to the description of the internal degrees of freedom of the hadrons. During the last decades, several approaches based on Regge theory and/or inspired on QCD have been proposed 
(See {\it e.g.} Refs. 
  \cite{Donnachie:1979yu,Cudell:2001pn,Block:2012ym,Avila:2002tk,Luna:2003kw,Broilo:2018els,Rasmussen:2018dgo,Fiore:2008tp,Luna:2008pp,Gotsman:1993ux,LHeureux:1985qwr,Margolis:1988ws,Block:1998hu,Durand:1988ax,Corsetti:1996wg,Fagundes:2013aja,Luna:2005nz,Broilo:2019xhs}), with reasonable success in the description of the pre -- LHC data \cite{pdg:2018} and the recent very precise data released by the ATLAS and TOTEM Collaborations \cite{Antchev:2013c,Antchev:2013b,Antchev:2015,Antchev:2016,Antchev:2019a,Antchev:2019b,Aad:2014,Aaboud:2016}. A comprehensive review of these different approaches for the treatment of the hadronic interactions was performed in Ref.~\cite{Pancheri:2016yel}.  As different approaches, based on very distinct assumptions, describe the current data for the energy dependence of the  total and elastic cross sections,  it is fundamental the theoretical and experimental analysis of  other observables that are more sensitive to the underlying physics and that allow us to advance in our  understanding about the strong interactions at high energies. The recent data for the $\rho$ parameter \cite{Antchev:2019a} and for the differential elastic cross section \cite{Antchev:2019c} are two examples of observables that have motivated an intense debate about the treatment of the hadronic interactions leading  several theoretical groups to modify its phenomenological models. Another way to probe the description of the hadronic interactions is the study of the diffractive excitation of the incident hadrons through the single and double diffractive processes, which are strongly sensitive to the modelling of the internal degrees of freedom of the hadrons. The analysis of these processes is the main goal of this paper.

In a collision of composite objects, one (or both) particle(s) is (are) excited to a higher mass state with the same quantum numbers, characterizing the single (double) diffractive processes. Therefore, the diffractive excitation arises from the fluctuating structure of the hadron. Many years ago, Good and Walker (GW) ~\cite{Good:1960} proposed to describe the diffractive excitation in terms of the eigenstates of the scattering operator, which are used to express the physical states. As the description of the eigenstates is directly related to the modelling of the proton wave function in the GW approach, we can expect that the study of single and double diffractive cross sections will be more sensitive to the underlying assumptions present in the phenomenological models.
A  model based on the GW approach was proposed by  Miettinen and Pumplin (MP) \cite{MP}, which have assumed that the  diffractive eigenstates correspond to QCD parton states. Recently, in Ref.~\cite{nosMP} our group have updated the MP model for $pp$ collisions and extended for proton -- nucleus interactions. In particular, it was demonstrated that  this model describes the current LHC data for the total, elastic and single diffractive cross sections. However, the double diffractive cross section was not estimated, since in the simplified approach developed in Ref.~\cite{nosMP} only the projectile was considered as having a substructure. Our goal in this paper is twofold. First, to present and discuss the implications of an alternative description of the diffractive eigenstates and for the modelling of the interaction between these eigenstates inspired in perturbative QCD. Second, to obtain predictions for the double diffractive cross sections that can be compared with future experimental LHC data. Moreover, we also will estimate the $\rho$ parameter and differential elastic cross section, which allow us to understand the current shortcomings of the model and point out the aspects that must be improved in the description of the hadronic structure.

The paper is organized as follows. In the next Section, we present a detailed derivation of the total, elastic, single and double diffractive cross section in the  Good -- Walker approach. A model for the description of the eigenstates and scattering amplitude is proposed  and the treatment of the average number of interactions in terms of the eikonal cross section is discussed. In Section \ref{sec:results} the  main parameters of the model are determined using the recent LHC data for the total cross section. Predictions for the elastic, total single diffractive and double diffractive cross sections are presented and compared with the current data. In addition, the descriptions of the $\rho$ parameter and differential elastic cross section are discussed. Finally, in Section \ref{sec:Sum}, our main conclusions are summarized.

%Studies on the model proposed by Lipari and Lusignoli in Phys.Rev.D \textbf{80}, 074014 (2009).

%Here we discuss the results present in the paper and propose some modifications.

\section{Formalism}

\subsection{General results}

Consider the collision between a projectile $P$  and a target $T$, which can be represented by the physical states $|P\rangle$ and $|T\rangle$, respectively. We will assume that both states have a substructure and  can be diffracted onto various particle states $\{|A\rangle\}$ and $\{|B\rangle\}$. Following the approach proposed by Good and Walker~\cite{Good:1960} many years ago, 
we will express the initial state in terms of the  eigenstates $\{|\Psi_i\rangle \}$ of the scattering operator $\hat{T}$, which form a complete set of normalized states. Consequently, we have that
\begin{equation}
 \vert I \rangle = \vert P, T\rangle = \sum_{ij} C^P_i C^T_j \vert \psi_i\psi_j\rangle
\end{equation}
and
\begin{eqnarray}
 \hat{T} \vert \psi_i \psi_j\rangle = t_{ij} \vert \psi_i \psi_j\rangle \,\,,
\end{eqnarray}
where the eigenvalues $t_{ij}$ depend on the particular configurations of the projectile and the target.
The final state system will be described by
\begin{eqnarray}
\vert F \rangle = \hat{T} \, \vert I \rangle = \sum_{i,j} C_i^P C_j^T t_{ij} \vert \psi_i\psi_j\rangle \,\,,
\end{eqnarray}
which implies
\begin{eqnarray}
\langle F | F \rangle = \sum_{i,j} |C_i^P|^2 |C_j^T|^2 |t_{ij}|^2 \,\,.
\end{eqnarray}
Identifying the quantities $P_i^P \equiv |C_i^P|^2$ and  $P_j^T \equiv|C_j^T|^2$ with the probability distributions for the configuration $i$ and $j$ in the projectile and target, respectively, we can write the above equation as follows:
\begin{eqnarray}
\langle F | F \rangle = \sum_{i,j} P_i^P  P_j^T |t_{ij}|^2 \equiv \langle t^2 \rangle_{P,T} \,\,,
\end{eqnarray}
where $\langle ... \rangle_{P,T}$ is the average over the configurations in the projectile {\it and} in the target.

Considering that the possible final  states are represented by $\{|A,B\rangle \}$, and that they form a complete set of eigenstates, we can write
\begin{eqnarray}
| F \rangle & = & \sum_{A,B} |A,B \rangle  \nonumber \\
& = & |P,T \rangle + \sum_{A \neq P} |A, T \rangle +
\sum_{B \neq T} |P, B \rangle + \sum_{A \neq P; B \neq T} |A, B \rangle \,\,.
\end{eqnarray}
Consequently, we also have that
\begin{eqnarray}
\langle F | F \rangle & = & \sum_{A,B} \langle F|A,B \rangle \langle A,B | F \rangle \nonumber  \\
& = & |\langle P, T | F \rangle |^2 + \sum_{A \neq P} |\langle A, T | F \rangle |^2 + \sum_{B \neq T} |\langle P, B | F \rangle |^2 +  \sum_{A \neq P; B \neq T} |\langle A, B | F \rangle |^2 \,\,.
\end{eqnarray}
On the other hand, using that
\begin{eqnarray}
\langle P, T | F \rangle & = & \sum_{i,j} |C_i^P|^2 |C_j^T|^2 t_{ij} \equiv \langle t \rangle_{P,T} \,\,,\\
\langle A, T | F \rangle|_{A \neq P} & = & \sum_{i,j} C_i^{*,A} C_i^P |C_j^T|^2 t_{ij} \,\,, \\
\langle P, B | F \rangle|_{B \neq T} & = & \sum_{i,j} |C_i^P|^2 C_j^{*,B} C_j^T  t_{ij}\,\,,\\
\langle A, B | F \rangle|_{A \neq P; B \neq T} & = & \sum_{i,j} C_i^{*,A} C_i^P C_j^{*,B} C_j^T  t_{ij}  t_{ij} \,\,,
\end{eqnarray}
we can write
\begin{eqnarray}
|\langle P, T | F \rangle |^2 + \sum_{A \neq P} |\langle A, T | F \rangle |^2 & = & \sum_{A} |\langle A, T | F \rangle |^2 = \sum_A \left| \sum_i C_i^{*,A} C_i^P \sum_j |C_j^T|^2 t_{ij}\right|^2 \nonumber \\
& = & \sum_A \left|\sum_i C_i^{*,A} C_i^P \langle t(j) \rangle_T \right|^2 = \sum_i C_i^{*,P} C_i^P \langle t(j) \rangle_T^2 = \langle \langle t \rangle_T^2 \rangle_P \,\,,
\end{eqnarray}
where we have used the completeness of the states 
$\{|A\rangle\}$, which implies $\sum_A C_i^{*,A} C_{i^{\prime}}^A = \delta_{i i^{\prime}}$. 
Similarly, we can derive that:
\begin{eqnarray}
|\langle P, T | F \rangle |^2 + \sum_{B \neq T} |\langle P, B | F \rangle |^2 & = & \sum_{B} |\langle P, B | F \rangle |^2 = \sum_B | \sum_j C_j^{*,B} C_i^T \sum_i |C_i^T|^2 t_{ij}|^2 \nonumber \\
& = & \sum_B \left|\sum_j C_j^{*,B} C_i^T \langle t(j) \rangle_P \right|^2 = \sum_j C_j^{*,T} C_j^T \langle t(j) \rangle_P^2 = \langle \langle t \rangle_P^2 \rangle_T \,\,,
\end{eqnarray}
and
\begin{eqnarray}
\sum_{A \neq P; B \neq T} |\langle A, B | F \rangle |^2 & = & 
\langle F | F \rangle -  |\langle P, T | F \rangle |^2 - \sum_{A \neq P} |\langle A, T | F \rangle |^2 - \sum_{B \neq T} |\langle P, B | F \rangle |^2  \nonumber \\
& = &  \langle t^2 \rangle_{P,T} - \langle \langle t \rangle_T^2 \rangle_P - \langle \langle t \rangle_P^2 \rangle_T  +  \langle t \rangle_{P,T}^2
\end{eqnarray}

Using the above relations we can define the associated cross sections in the impact parameter space as follows:
\begin{itemize}
    \item The elastic cross section:
\begin{eqnarray}
\frac{d^2 \sigma_{el}}{d^2b} = |\langle P, T | F \rangle |^2 =  \langle t \rangle_{P,T}^2 \,\,
\label{eq:elas}
\end{eqnarray}
    \item The  cross section for the single diffractive excitation of the projectile:
\begin{eqnarray}
\frac{d^2 \sigma_{SD}^P}{d^2b} = \sum_{A \neq P} |\langle A, T | F \rangle |^2 = \langle \langle t \rangle_T^2 \rangle_P - \langle t \rangle_{P,T}^2 \,\,; 
\label{eq:sdp}
\end{eqnarray}
    \item The  cross section for the single diffractive excitation of the target:
\begin{eqnarray}
\frac{d^2 \sigma_{SD}^T}{d^2b} = \sum_{B \neq T} |\langle P, B | F \rangle |^2 = \langle \langle t \rangle_P^2 \rangle_T - \langle t \rangle_{P,T}^2 \,\,;
\label{eq:sdt}
\end{eqnarray}
\item The double diffractive cross section:
\begin{eqnarray}
\frac{d^2 \sigma_{DD}}{d^2b} = 
\sum_{A \neq P; B \neq T} |\langle A, B | F \rangle |^2 & =  &  \langle t^2 \rangle_{P,T} - \langle \langle t \rangle_T^2 \rangle_P - \langle \langle t \rangle_P^2 \rangle_T  +  \langle t \rangle_{P,T}^2 \,\,.
\label{eq:sdd}
\end{eqnarray}
\end{itemize}
In addition, we can define the total single diffractive cross section by:
\begin{eqnarray}
\frac{d^2 \sigma_{SD}}{d^2b} = \frac{d^2 \sigma_{SD}^P}{d^2b} + \frac{d^2 \sigma_{SD}^T}{d^2b} = \langle \langle t \rangle_T^2 \rangle_P + 
\langle \langle t \rangle_P^2 \rangle_T - 2 \langle t \rangle_{P,T}^2 \,\,,
\label{eq:sdtot}
\end{eqnarray}
and the total diffractive cross section:
\begin{eqnarray}
\frac{d^2 \sigma_{diff}}{d^2b} = \frac{d^2 \sigma_{SD}}{d^2b} + \frac{d^2 \sigma_{DD}}{d^2b} =  \langle t^2 \rangle_{P,T} - \langle t \rangle_{P,T}^2 \,\,.
\label{eq:diftot}
\end{eqnarray}
Finally, using the optical theorem, the total cross section is given by
\begin{eqnarray}
\frac{d^2 \sigma_{tot}}{d^2b} = 2 \langle t \rangle_{P,T} \,\,.
\label{eq:tot}
\end{eqnarray}

The final results obtained in this subsection, Eqs.~(\ref{eq:elas}) -- (\ref{eq:tot}),    already have appeared  in several papers in the literature, considering different models for the description of the eigenstates and for the scattering amplitude (See {\it e.g.}. Refs. \cite{lund,Ryskin:2007qx,levin}). Due to the generality of the results and for completeness of our study, we decided to  present a detailed derivation of the cross sections. In the next subsection, we will discuss our model for the treatment of the average over configurations of the projectile and target as well as  for the description of the scattering amplitude for the interaction between these configurations.

\subsection{A model for the eigenstates and scattering amplitude }

One of the most important questions in the description of the hadronic interactions using the Good -- Walker approach is which are the diffractive eigenstates $\{|\Psi_i\rangle\}$. During the last decades, several authors have considered different approaches to treat this aspect \cite{MP,Ryskin:2007qx, thoma,lipari,Chou:1968bc,Chou:1968bg,Bialas:1972cx,Bialas:1972ru,Fialkowski:1975ta,VanHove:1976fi,VanHove:1977kc,Nussinov:1979ca, Bertsch:1981py,Alberi:1981sz,DiasdeDeus:1987yw,Margolis:1988ws,Durand:1988ax,Heiselberg:1991is,Fletcher:1992sy}. For instance, in Ref. \cite{MP}  
Miettinen and Pumplin (MP)  assumed that these eigenstates correspond to QCD   parton states, which can come on shell through interaction with the target. Last year, such model was updated and extended for proton -- nucleus collisions by our group, with the predictions describing the LHC data \cite{nosMP}. In recent years, in a series of papers, G. Gustafson, L. Lonnbland and collaborators have proposed to assume the  color quark -- antiquark dipoles as the diffractive eigenstates \cite{lund}, with the  energy evolution and interaction between these dipoles being described in terms of the  Mueller's dipole model \cite{mueller}. Such promising approach is currently being updated and improved \cite{cristian}. Here we will consider an alternative description of the diffractive eigenstates and for the interaction between the different configurations of the hadronic structure. 
Inspired by the study performed in Ref.~\cite{lipari}, we will assume that the distinct configurations $\bbC_i$ can be represented by a continuum  distribution, with each configuration  having a probability $P_{hi}(\bbC_i)$. Consequently, we can perform the following identifications
\begin{align}
 \sum_i |C_i^P|^2 & \to \int d\bbC_1 P_{h1}(\bbC_1) \text{ for the projectile,}\\
 \sum_i |C_i^T|^2 & \to \int d\bbC_2 P_{h2}(\bbC_2) \text{ for the target}
\end{align}

\noindent and
\begin{equation}
 t_{ij}(b,s) \to t(b,s,\bbC_1,\bbC_2) \,\,,
\end{equation}
where the impact parameter and energy dependencies of the scattering amplitude between the distinct configurations is explicitly shown. Considering that the incident hadrons are constituted by partons and that for a given combination of configurations $\bbC_1$ and $\bbC_2$ the  expected number of interactions between these partons in a collision with impact parameter $b$ can be represented by 
${n(b,s,\bbC_1,\bbC_2)}$, and then $t(b,s,\bbC_1,\bbC_2) \propto n(b,s,\bbC_1,\bbC_2)$ at leading order. In order to take into account of the multiple interactions between the configurations we will assume that  $t(b,s,\bbC_1,\bbC_2)$ is given by the eikonal form
\begin{equation}
 t(b,s,\bbC_1,\bbC_2) = 1-\exp\left[-\frac{n(b,s,\bbC_1,\bbC_2)}{2}\right]\,\,.
\end{equation}
In addition, we will assume that the distribution of parton configurations is independent of the impact parameter, which implies that we can write
\begin{equation}
 n(b,s,\bbC_1,\bbC_2) = \langle n(b,s)\rangle \alpha(\bbC_1)\alpha(\bbC_2),
\end{equation}
where $\langle n(b,s)\rangle$ is the average number of interactions at a fixed $b$ and center -- of -- mass energy $\sqrt{s}$ and the functions $\alpha(\bbC_i)$  depend on the configurations of the incident hadrons.
 We have therefore
\begin{equation}
 \int d\bbC_1 \int d\bbC_2 P_{h1}(\bbC_1)P_{h2}(\bbC_2) e^{-n(b,s,\bbC_1,\bbC_2)/2} = \int_0^\infty d\alpha_1\int_0^\infty d\alpha_2 p(\alpha_1)p(\alpha_2) e^{-\langle n(b,s)\rangle \alpha_1\alpha_2/2},
\end{equation}
where we have defined the auxiliary functions $p(\alpha_i)$ by
\begin{equation}
 p(\alpha_i) = \int d\bbC_i P_{hi}(\bbC_i)\delta[\alpha(\bbC_i) - \alpha_i],
\end{equation}
which satisfy the following constraints:
\begin{equation}
 \int_0^\infty d\alpha_i p(\alpha_i) = 1.
 \label{eq:norm_p_alpha}
\end{equation}
and 
\begin{equation}
 \int_0^\infty d\alpha_i \alpha_i p(\alpha_i)   = 1 \quad (i=1,\,2) \,\,.
 \label{eq:average_p_alpha}
\end{equation}
Consequently, the average over configurations present in the Eqs. (\ref{eq:elas}) -- (\ref{eq:tot}) will be given by % 
\begin{itemize}
 \item Average over the projectile configurations: $\langle t^n \rangle_P = \int_0^\infty d\alpha_1 p(\alpha_1) t^n(b,s,\alpha_1,\alpha_2)$;
 \item Average over the target configurations: $\langle t^n \rangle_T = \int_0^\infty d\alpha_2 p(\alpha_2) t^n(b,s,\alpha_1,\alpha_2)$;
 \item Average over projectile and target configurations:  $\langle t^n \rangle_{PT} = \int_0^\infty d\alpha_1 \int_0^\infty d\alpha_2 p(\alpha_1) p(\alpha_2) t^n(b,s,\alpha_1,\alpha_2)$.
\end{itemize}

% 
% NEW
% 
{ The form of the probability distribution $p(\alpha_i)$, that describes the fluctuations of the hadron configurations, is still an open problem.  However, we expect that this distribution presents  the following properties: it must be defined for positive values of its variable ($\alpha$) and it has to show the expected limit, $p(\alpha)\to\delta(\alpha-1)$, when its variance goes to zero, corresponding to no-fluctuations. Additionally, it is interesting to have an analytical structure that allows us, in some extent, to obtain analytical (closed) expressions. The gamma distribution{, with variance $w$,}  
\begin{equation}
  p(\alpha_i) = \frac{1}{w\Gamma(1/w)}\left(\frac{\alpha_i}{w}\right)^{-1+1/w}e^{-\alpha_i/w}\,\, ,
  \label{eq:p_alpha}
\end{equation}

\noindent also used in other analysis \cite{thoma,lipari}, has the above properties and we shall consider it in this work.
Eq.~\eqref{eq:p_alpha} satisfies the constraints of Eqs.~\eqref{eq:norm_p_alpha} and \eqref{eq:average_p_alpha}. Moreover, for $w\to 0$ we have $p(\alpha) \to \delta(\alpha-1)$, corresponding to the case of no-fluctuations which implies no dissociative proccess.
For simplicity, we will assume that the variance $w$ of the distribution is independent of $i$ since we are considering the collision of identical hadrons.
}
% 
% 
% 
% Following Ref.~\cite{thoma} we will assume that the functions $p(\alpha_i)$ can be represented by  gamma distributions. For simplicity, we will assume that the variance  $w$ of the distribution is independent of $i$ since we are considering the collision of identical hadrons, {\it i.e.} we consider that
% % 
% \begin{equation}
%   p(\alpha_i) = \frac{1}{w\Gamma(1/w)}\left(\frac{\alpha_i}{w}\right)^{-1+1/w}e^{-\alpha_i/w}\,\,.
% \end{equation}
Such assumption allows to calculate the average over configurations necessary to calculate the total, elastic, total single diffractive and double diffractive cross sections, which will be given by
\begin{align}
\langle t \rangle_{PT} & = 1- x^{1/w}U(1/w,1,x),\\
 \langle \langle t \rangle_T^2\rangle_{P} & = 1 - 2x^{1/w}U(1/w,1,x) + x^{1/w}U(1/w,1-1/w,x) =  \langle \langle t \rangle_P^2\rangle_{T}, \\
 \langle t^2\rangle_{PT} & = 1 - 2x^{1/w}U(1/w,1,x) + (x/2)^{1/w}U(1/w,1,x/2),
\end{align}

\noindent where 
\begin{equation}
 x \equiv \frac{2}{\langle n(b,s) \rangle w^2}
\end{equation}

\noindent and
\begin{equation}
 U(a,b,x) = \frac{1}{\Gamma(a)}\int_0^\infty e^{-xu} u^{a-1}(1+u)^{b-a-1} du
\end{equation}

\noindent is the confluent hypergeometric function \cite{abramowitz}. With these results, all cross sections can be estimated once we assume a model for the average number of partonic interactions $\langle n(b,s) \rangle$.

% \section{$\sigma_{eik}$ and fits to data}
\subsection{The average number of interactions}

% The treatment of the average number of interactions is still an open question in the literature. In particular, the factorization or not of the impact parameter and energy dependencies. For simplicity, we will assume the {\it Ansatz} proposed by Durand and Pi many years ago \cite{Durand:1988ax}, in which these two dependencies are factorized as follows $\langle n(s,b) \rangle~=~\sigma_\text{eik}(s) A(b)$, with the energy dependence being described by the eikonal cross section $\sigma_\text{eik}$ and the impact parameter dependence by geometric factor $A(b)$ that describes the overlap of hadronic matter in the collision. Inspired by Ref.~\cite{Durand:1988ax}, we will assume that

% 
% NEW 
% 
{
The treatment of the average number of interactions $\langle n(s,b) \rangle$ is still an open question in the literature. 
{
However, it is expect that this quantity depends on how the partons are distributed inside the hadron in the $b$-plane and on the intensity of the parton-parton interactions. Collisions at different impact parameters will result in a distinct number of interactions, with the amount of them depending on the number of partons in each of the interacting hadrons. In principle, the partons distribution in $b$-plane can be modelled by the overlap of hadronic matter in the collision $A(s,b)$, which is related to the Fourier transform of the hadron form factor. On the other hand, the intensity is expected to be determined by the dynamics of the interaction, {\it i.e.} by the cross section $\sigma_{eik}(s,b)$ that describes the interaction between the incident hadrons for a given  center of mass energy $\sqrt{s}$ and impact parameter $b$.
For simplicity, in our analysis we will consider the \textit{Anzats} proposed by Durand and Pi \cite{Durand:1988ax}, and considered in several analysis \cite{Margolis:1988ws,Block:1998hu,Luna:2005nz,Broilo:2019xhs,Fagundes:2017xli}, which assumes that $A(s,b)$ depends only on $b$, while $\sigma_{eik}(s,b)$ is a function of energy only.
 Therefore, it is assumed that $\langle n(s,b) \rangle = \sigma_{eik}(s)A(b)$. We note that the factorization or not of the impact parameter and energy dependencies in the description of average number of interactions is one important aspect that can be improved in the future, considering {\it e.g.} the description of the cross section $\sigma(s,b)$ in terms of the generalized parton distributions, 
which provide  information about how partons are distributed in the plane transverse to the direction in which the hadron is moving (For a review see, e.g. \cite{diehl}). Consequently, inspired by Ref.~\cite{Durand:1988ax}, in our analysis we will assume that }
%also an open problem. For simplicity, in this work we will consider it valid. Inspired by Ref.~\cite{Durand:1988ax}, we have
%
%The partons distribution in $b$-plane is given by the geometric factor $A(s,b)$, that describes the overlap of hadronic matter in the collision. The overlap function is also related to the Fourier transform of the hadron form factor. On the other hand, the intensity is determined by the dynamics of the interaction and is described by the eikonal cross section $\sigma_{eik}(s,b)$. 
% 
%However, we consider that the number of interactions can be affected by the partons distribution inside the hadron in the $b$-plane and by the intensity of the parton-parton interactions. The partons distribution in $b$-plane is described by the overlap of hadronic matter in the collision $A(s,b)$ (related to the Fourier transform of the hadron form factor). On the other hand, the intensity is determined by the dynamics of the interaction and is described by the eikonal cross section $\sigma_{eik}(s,b)$.
%In the \textit{Anzats} proposed by Durand and Pi \cite{Durand:1988ax}, and considered in several analysis \cite{Margolis:1988ws,Block:1998hu,Luna:2005nz,Broilo:2019xhs,Fagundes:2017xli}, we assume that $A(s,b)$ depends only on $b$, while $\sigma_{eik}(s,b)$ is a function of energy only. 
\begin{equation}
 \langle n(s,b) \rangle = \sigma_\text{eik}(s) \frac{b^3}{96\pi r_0^5} K_3(b/r_0),
 \label{eq:factb}
\end{equation}
with $r_0 = 1.0$ GeV$^{-1}$ for simplicity and where the modified Bessel function of the second kind $K_3(b/r_0)$ results from the Fourier transform of the dipole form factor assumed for the proton. 
} {Another important open question is the description of $\sigma_\text{eik}(s)$ which receives contributions from the perturbative and nonperturbative regimes of QCD. The factorization of these contributions in the total cross sections is not proven, being dependent on the model assumed to describe the hadronic interaction. However, it is expected that at large values of $\sqrt{s}$ the energy dependence will be driven by  hard partonic interactions, which can be described by perturbative QCD. Following previous  QCD-based models, we will assume  that soft and hard contributions for the description of a hadronic collision are additive, which implies that  
%the description of $pp$ and $\bar{p}p$ scatterings  are additive with respect to soft and a hard partonic interactions in the hadron-hadron collision. With this assumption in mind e will consider that 
$\sigma_\text{eik}$ can be expressed as a combination of hard and soft contributions}
\begin{equation}
 \sigma_\text{eik} (s) =  \sigma_\text{pQCD}(s) + \sigma_\text{soft}(s)\,\,,
 \label{eiksec}
\end{equation}
where $\sigma_\text{pQCD}(s)$ is expected to determine the high energy behavior of $\langle n(s,b) \rangle$, while 
$\sigma_\text{soft}(s)$ determines the associated behavior at low energies. In our analysis, we will assume that the hard contribution can be expressed in terms of the minijet cross section as follows:
\begin{eqnarray}
\sigma_\text{pQCD} (s) & = & {\cal{K}} \sigma_\text{minijet}(s) \nonumber \\
&=&\sum_{i,j=q,\bar{q},g}\,\frac{{\cal{K}}}{1+\delta_{ij}}\,\int^{1}_{0} dx_{1}\int^{1}_{0} dx_{2}\int^{\infty}_{Q^{2}_{min}}\!\!\!\!d\vert\hat{t}\vert\,\frac{d\hat{\sigma}_{ij}}{d\vert\hat{t}\vert}(\hat{s},\hat{t})
 \times  f_{i/P}(x_{1},\vert\hat{t}\vert)\,f_{j/T}(x_{2},\vert\hat{t}\vert)\,\Theta\left(\frac{\hat{s}}{2}-\vert \hat{t} \vert\right)\,\,,
\label{sigQCD}
\end{eqnarray}
where the factor ${\cal{K}}$ takes into account of next -- to -- leading order corrections to the minijet cross section  \cite{Sarcevic:1988tu}, $x_1$ and $x_2$ are the momentum fractions of the partons inside of hadrons  $P$ and $T$, 
$\hat{s}$ and $\hat{t}$ are the Mandelstam variables for the partonic collision, $d\hat{\sigma}_{ij}/d\vert \hat{t}\vert $ is the differential cross section for $ij$ scattering, and $f_{i/h}$ are the parton distribution functions of the hadron $h$. Moreover,  the integration over $\vert \hat{t}\vert$  satisfy the physical condition $\hat{s}>2\vert \hat{t}\vert>2Q^{2}_{min}$, where $\hat{s} = x_{1}x_{2}s$ and $Q^{2}_{min}$ is the minimal momentum transfer in the hard scattering. Since the gluon distribution becomes dominant in the high -- energy regime, we will include in our calculations all process  with at least one gluon in the initial state. 
We will assume that the parton distribution functions can be described by the post -- LHC fine -- tunned parametrization proposed by the CTEQ-TEA group in  Ref.~\cite{CT14}. Moreover, the infrared divergences present in the elementary subprocesses at low transferred momenta will be regularized using the approach proposed in the Refs.~\cite{Luna:2005nz,Broilo:2019xhs}.

%{\bf ([Mateus] The partonic scattering amplitudes are written by means of the standard QCD cross section in Eq.~(\ref{sigQCD}) convoluted with partonic distribution functions (PDFs).  It is well known that at the limit of low transferred momenta these elementary processes are plagued by infrared divergences and thus must be regularized through out some cutoff procedure. One natural way to overcome this problem was introduced a few decades ago by Cornwall \cite{Cornwall:1981zr} and is based on the increasing evidence that the gluon may develop a momentum -- dependent mass. This dynamical mass introduces a natural scale able to separate the perturbative from the nonperturbative QCD region and allow us to explore the nonperturbative character of QCD in order to describe hadronic observables. For a recent treatment about this topic see Refs.~\cite{Luna:2005nz,Bahia:2015hha,Broilo:2019xhs,Broilo:2019yuo}. In our calculations we will investigate the effects of an updated set of partonic distribution functions post -- LHC fine -- tunned parametrization proposed by the CTEQ-TEA group in  Ref.~\cite{CT14}.)}

%assume that the parton distribution functions can be described by the post -- LHC fine -- tunned parametrization proposed by the CTEQ-TEA group in  Ref.~\cite{CT14}.)}

On the other hand, motivated by the Regge-Gribov phenomenology, we will consider that the soft contribution is given by
\begin{align}
 \sigma_\text{soft} & = A_1 \left(s/s_0\right)^{-\delta_1} \pm A_2 \left(s/s_0\right)^{-\delta_2} + \sigma_0\,\,,
\end{align}
where  the first and second terms correspond to even and odd Reggeon exchange, respectively, and the constant term $\sigma_0$ to the critical Pomeron exchange. In our study, we will assume  $s_0 = 25$~GeV$^2$. It is important to mention that in the present model, we shall interpret these Reggeons as effective contributions since they do not appear directly at the Born level of the amplitude. The soft term also allow us to describe simultaneously crossed reactions, with  the plus sign corresponding to $\bar{p}p$ scattering and the minus sign to $pp$. The parameters appearing in the soft cross section, $A_1$, $A_2$, $\delta_1$, $\delta_2$ and $\sigma_0$ will be determined from fits to experimental data.

\section{Results}
\label{sec:results}

%\subsection{Dataset and fit procedures}

In order to determine the parameters of the model, $w$, $A_1$, $A_2$, $\delta_1$, $\delta_2$ and $\sigma_0$ we consider fits to experimental data for the $pp$ and $\bar{p}p$ total cross sections. 
% 
% NEW 
% 
{ Note that $\sigma_{tot}$ is related to the average of the eigenvalues [Eq.~\eqref{eq:tot}] and, consequently, less sensitive to $w$, which determines $p(\alpha)$, than single diffractive cross section, for example. On the other hand, the data of $\sigma_{tot}$ consists of one of the best dataset available, considering the large energy range covered and the precision of the data, especially at high energies. 
Moreover, as  the current precision of the experimental data for $\sigmasd$ is still small, we have chosen not include it in our fit. However, it is important to emphasize that we have verified that the description of $\sigmatot$ is sensitive to the value of $w$.}
{ Therefore, our dataset comprise data on $\sigma_{tot}$ obtained only in } accelerator experiments { covering} the energy range from 5~GeV to 13~TeV. Data below 1.8~TeV were obtained from the Particle Data Group \cite{pdg:2018}, while for the LHC energies, we included data obtained by TOTEM Collaboration in the energies of 2.76, 7, 8 and 13 TeV \cite{Antchev:2013c,Antchev:2013b,Antchev:2015,Antchev:2016,Antchev:2019a,Antchev:2019b} and data by ATLAS Collaboration at 7 and 8~TeV \cite{Aad:2014,Aaboud:2016}. For all data, we considered statistical and systematical uncertainties added in quadrature. The only fixed parameter is ${\cal{K}}$, which were considered three possible values: 1.5, 2.0 and 2.5. We believe that the chosen values correspond to reasonable choices given that this parameter effectively takes into account corrections of higher orders. Therefore, our model has six free parameters, being five of them associated to the description of the soft cross section. All fits were performed using the class TMinuit from ROOT Framework \cite{tminuit} through the MIGRAD algorithm. We consider a $\chi^{2}$ fitting procedure where the data reduction assumes an interval $\chi^{2}-\chi^{2}_{min}=7.04$ corresponding, in the case of normal errors, to the projection of the $\chi^{2}$ hypersurface containing $68.3\%$ of probability, namely uncertainties in the free parameters with 1$\sigma$ of confidence level.
% All fits were done using the class TMinuit from ROOT Framework \cite{tminuit}. Uncertainties in the free parameters corresponds to 1$\sigma$ ($\sim 68\%$) of confidence level.

\begin{table}[t]
 \centering
\begin{tabular}{c|c|c|c}\hline\hline
 Parameters     & \multicolumn{3}{c}{${\cal{K}}$ fixed}\\\hline
 ${\cal{K}}$            & 1.5 (fixed)      & 2.0 (fixed)       & 2.5 (fixed)       \\
 \hline
 $w$            & 1.363 $\pm$ 0.050& 1.720 $\pm$ 0.054 & 2.007 $\pm$ 0.054 \\
 $A_1$ (mb)     & 42.9 $\pm$ 3.9   & 54.6 $\pm$ 4.9    & 65.9 $\pm$ 5.9    \\
 $\delta_1$     & 1.42 $\pm$ 0.11  & 1.39 $\pm$ 0.11   & 1.37 $\pm$ 0.10   \\
 $A_2$ (mb)     & 37.4  $\pm$ 2.7  & 47.5 $\pm$ 3.4    & 57.1 $\pm$ 4.1    \\
 $\delta_2$     & 0.585 $\pm$ 0.044& 0.588 $\pm$ 0.045 & 0.591 $\pm$ 0.046 \\
 $\sigma_0$ (mb)& 122.5 $\pm$ 3.5  & 146.6 $\pm$ 4.4   & 168.9 $\pm$ 5.0   \\\hline
 $\chi^2/$dof   & 3.56             & 3.31              & 3.12              \\
 dof            & 171              & 171               & 171               \\\hline\hline
 \end{tabular}
 \caption{\label{tab1}Results of fits to $\sigmatot$ data from $pp$ and $\bar{p}p$ scattering with ${\cal{K}}$ fixed in three values: 1.5, 2.0 and 2.5. Information on reduced $\chi^2$ and degrees of freedom is also presented.}
 \end{table}

\begin{figure}[t]
 \centering
 \includegraphics[scale=0.54]{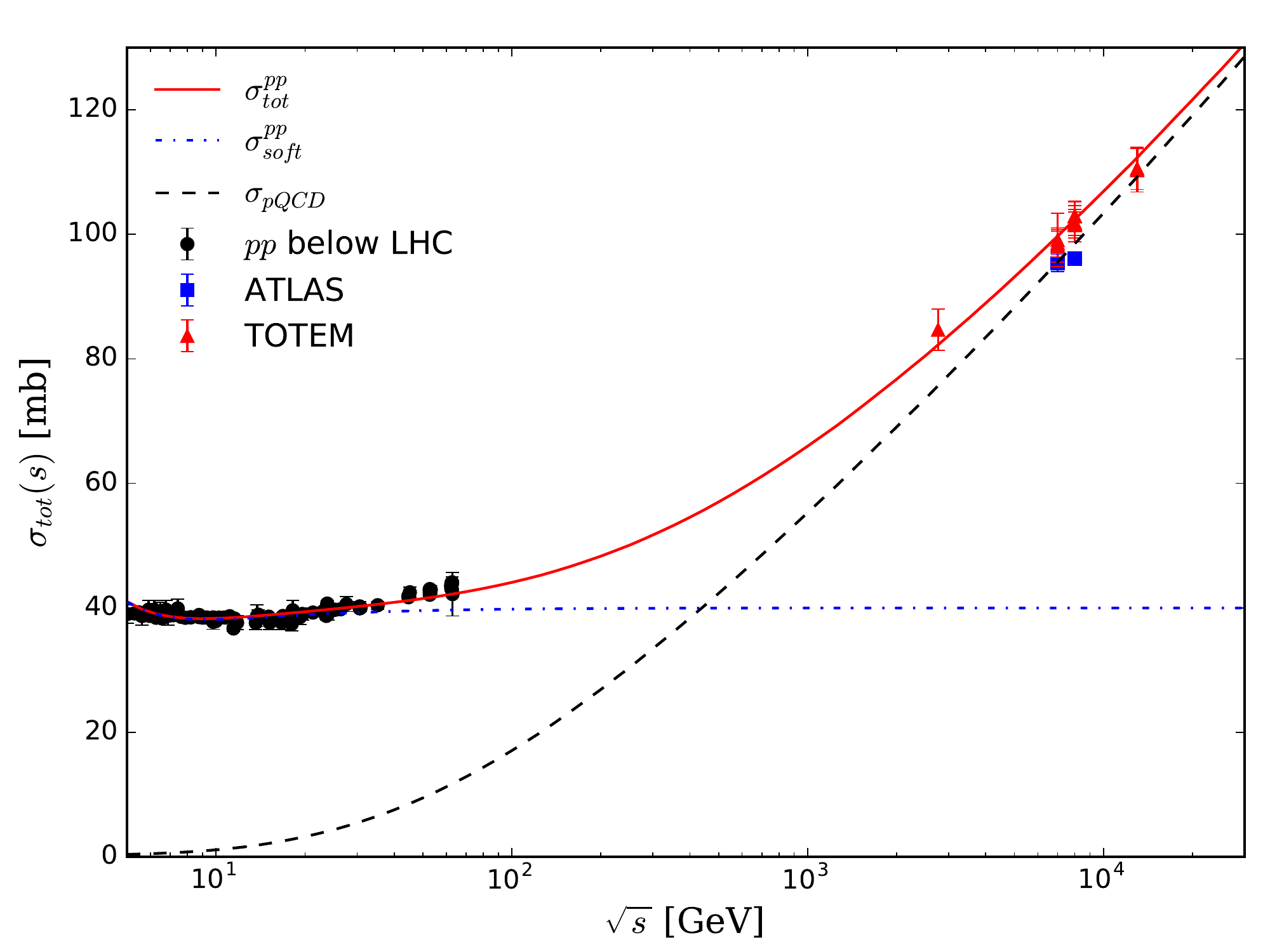}
  \caption{\label{figAdd} { Hard and soft contributions for the total $pp$ cross section (${\cal K}$=2.0). The blue dot-dashed line represents the result derived assuming that $\sigma_{eik}$ is given  only by the soft component, while  for the black dashed line it is assumed that $\sigma_{eik} = \sigma_{pQCD}$. Finally, for the red solid line we have that $\sigma_{eik}$ is given by the sum of the hard and soft components.}}

\end{figure}

\begin{figure}[t]
 \centering
 \includegraphics[scale=0.74]{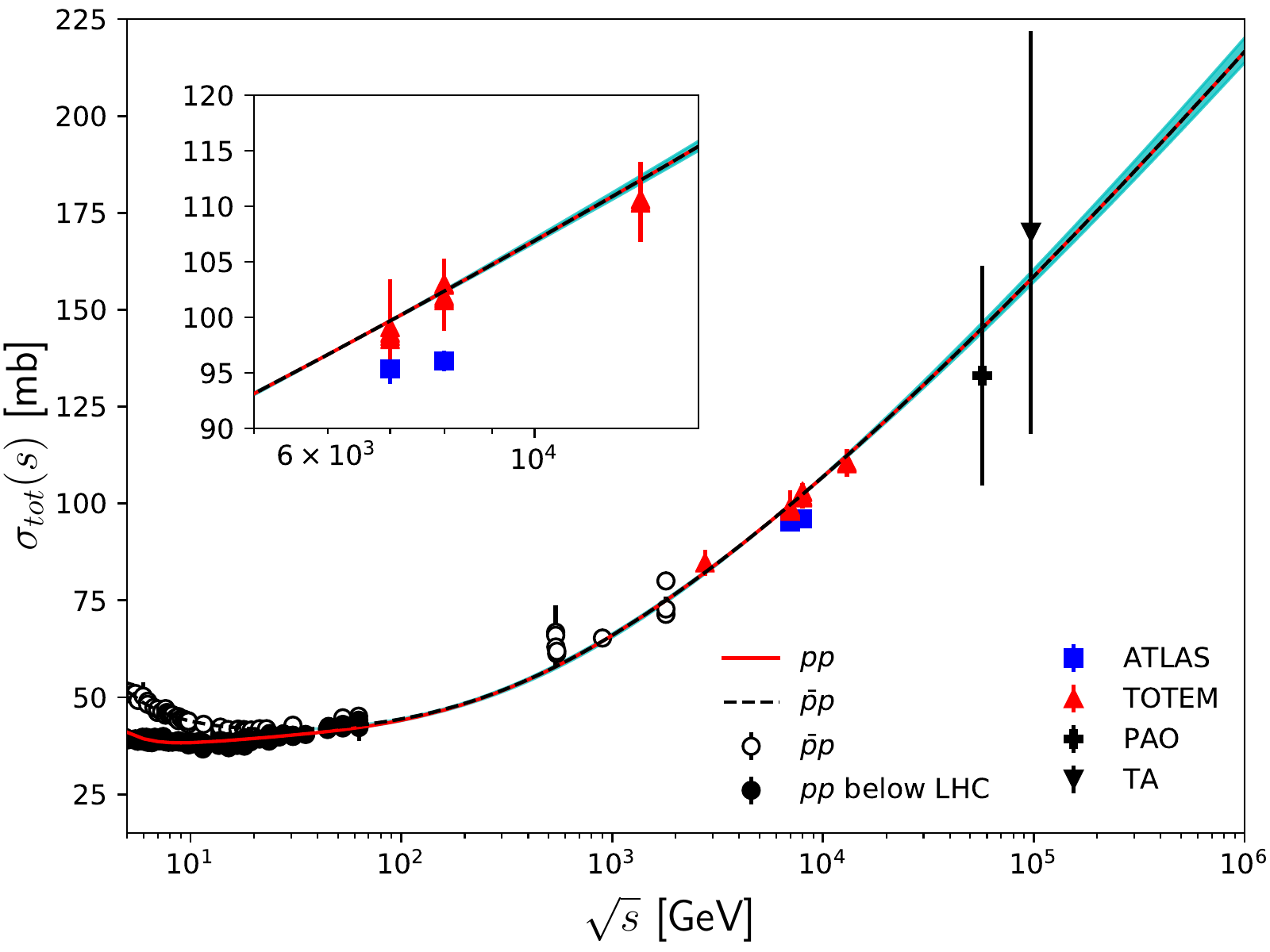}
  \caption{\label{fig1} Result of fit to $\sigmatot$ data from $pp$ and $\bar{p}p$ scattering (including TOTEM and ATLAS data). The central curve corresponds to the case with ${\cal{K}}=2$ while the cyan band around it indicates the allowed region between the results obtained with ${\cal{K}}=1.5$ and ${\cal{K}}=2.5$. 
  {Cosmic-ray data from the Pierre Auger Observatory (PAO) and from Telescope Array (TA) were not included in the fits, but are displayed in the figure for comparison.}
  Data from Refs. \cite{pdg:2018,Antchev:2013c,Antchev:2013b,Antchev:2015,Antchev:2016,Antchev:2019a,Antchev:2019b,Aad:2014,Aaboud:2016,Abreu:2012,Abbasi:2015}.}

\end{figure}

The results of the fits for the different values of ${\cal{K}}$ are presented  in Table~\ref{tab1}. { In order to estimate the impact of the hard and soft components of $\sigma_{eik}$ in the description of $\sigmatot$, in  Fig.\ref{figAdd} we present our results for the  $pp$ total cross section assuming that $\sigma_{eik}$ is either given by the soft component, {\it i.e.} $\sigma_{eik} = \sigma_{soft}$, or as $\sigma_{eik} = \sigma_{pQCD}$. Bearing in mind that soft, nonperturbative, physics plays a crucial part in the description of elastic, total and diffractive cross sections, we see that the high energy behavior is driven mainly by the partonic hard sector, being sensitive to the assumptions assumed in its calculation. The comparison with the experimental data for $pp$ and $p\bar{p}$ collisions} is shown in Fig.~\ref{fig1}. The central curve corresponds to the case with ${\cal{K}}=2$ while  the band around it indicates the allowed region between the results obtained with ${\cal{K}}=1.5$ and ${\cal{K}}=2.5$. {The analysis for different values of ${\cal{K}}$ is motivated by  the importance of the $\sigma_{pQCD}$ contribution at high energies.}
 We have verified  that the $\chi^2/$dof decreases by $\approx$ 20 \% if the ATLAS data is not included in the analysis. The results presented in Fig.~\ref{fig1} demonstrate that our model is able to describe the data for the total cross section and that the impact of different values of ${\cal{K}}$ is small in this observable.

\begin{figure}[t]
 \centering
 \includegraphics[scale=0.54]{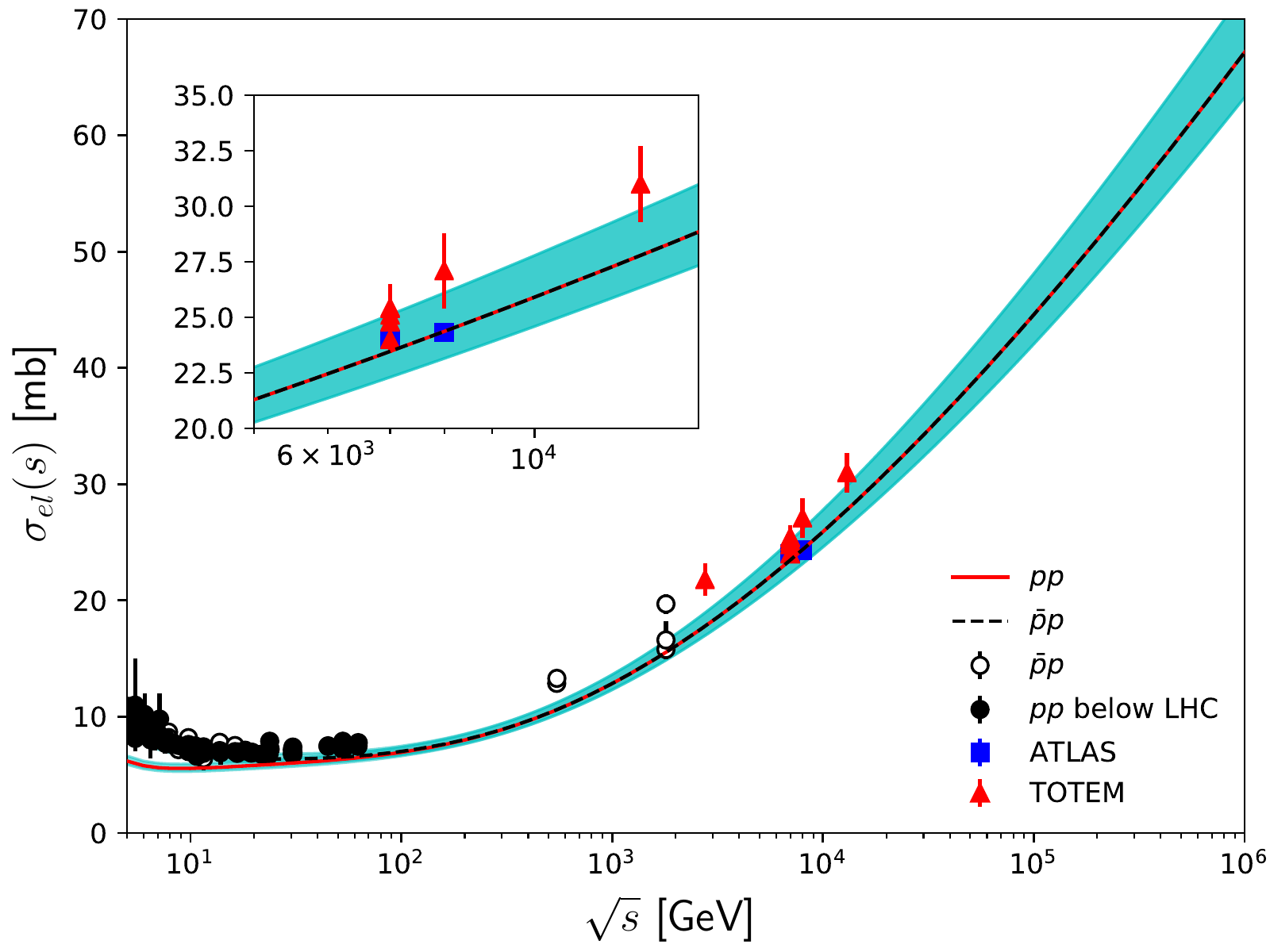}
  \includegraphics[scale=0.54]{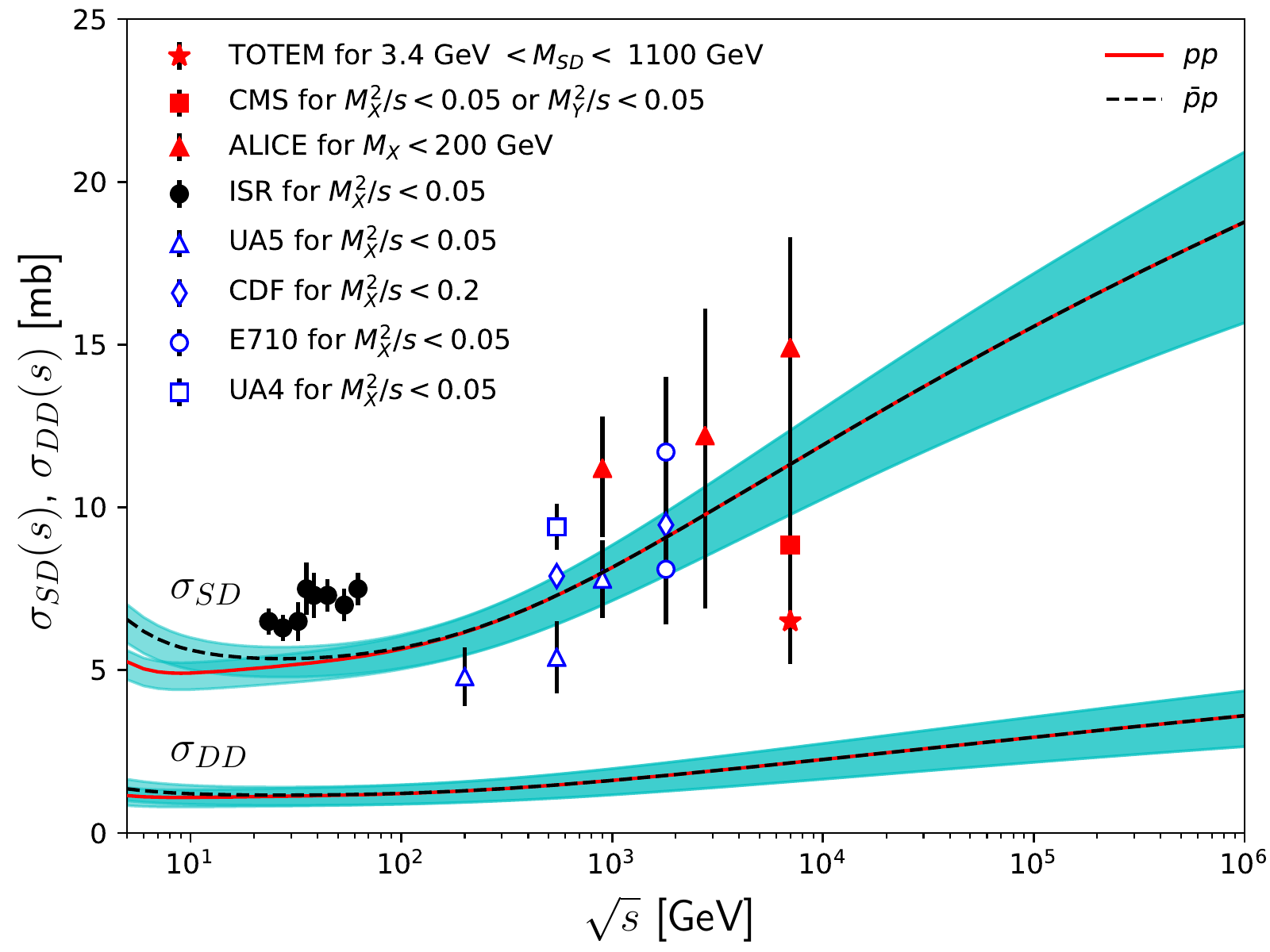}
  \caption{ Predictions for the elastic (left panel) and total single diffractive and double diffractive (right panel) cross sections. The central curve corresponds to the case with ${\cal{K}}=2$ while the cyan band around it indicates the allowed region between the results obtained with ${\cal{K}}=1.5$ and ${\cal{K}}=2.5$. Data for elastic cross section from \cite{pdg:2018,Antchev:2013c,Antchev:2013b,Antchev:2015,Antchev:2019b,Nemes_talk:2018} and single diffractive cross section from \cite{Armitage:1982,Cartiglia:2013,alice:2013,cms:2015,ua5:1986a,Alner:1987,Bernard:1987,Amos:1990,Amos:1993,cdf:1994a}.}
  \label{fig:sigelas}
\end{figure}

As all parameters of the model have been constrained by the experimental data for the total cross section, we can derive parameter free predictions for the elastic, total single diffractive and double diffractive cross sections. The predictions are presented in Fig.~\ref{fig:sigelas} and compared with the $pp$ and $p\bar{p}$ data. We can see that the model describes quite well the data at high energies. The only exception is the SD data at low energies, which we are not able to describe the normalization. Such result can be an indication that the naive {\it Ansatz} for the soft cross section must be improved. Another important aspect that we would like to emphasize is that  the predictions are more sensitive to the value of the ${\cal{K}}$ factor, in particular, for large energies. The comparison with the elastic data indicates that a larger value of ${\cal{K}}$ implies a better description. Regarding to the double diffractive cross section, we predict a mild increasing with the energy in the region probed by the LHC. Future experimental data for this observable will be very useful to probe the model proposed  in this paper to describe the diffractive excitation in hadronic collisions.

\begin{figure}[t]
 \centering
  \includegraphics[scale=0.385]{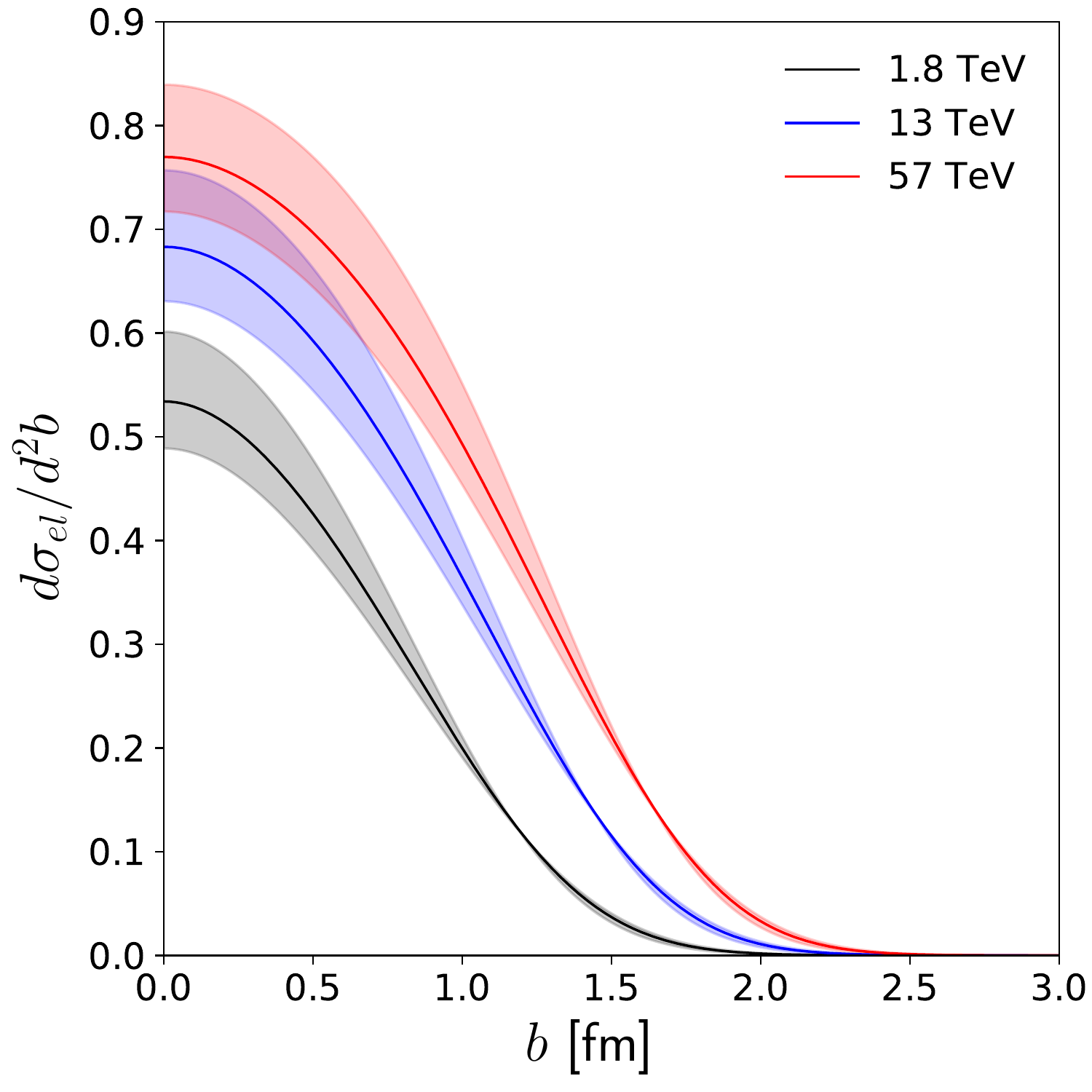}
  \includegraphics[scale=0.385]{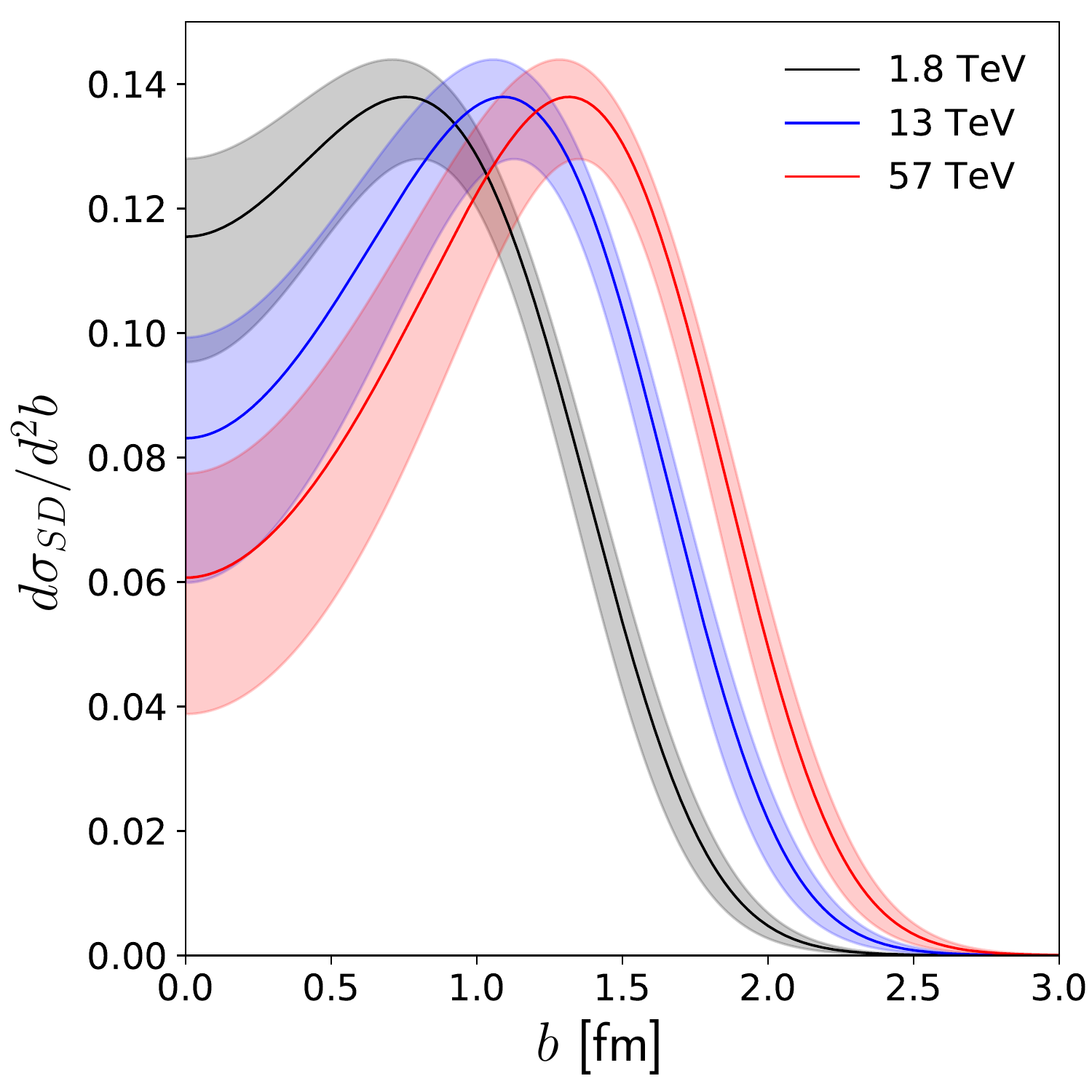}
  \includegraphics[scale=0.385]{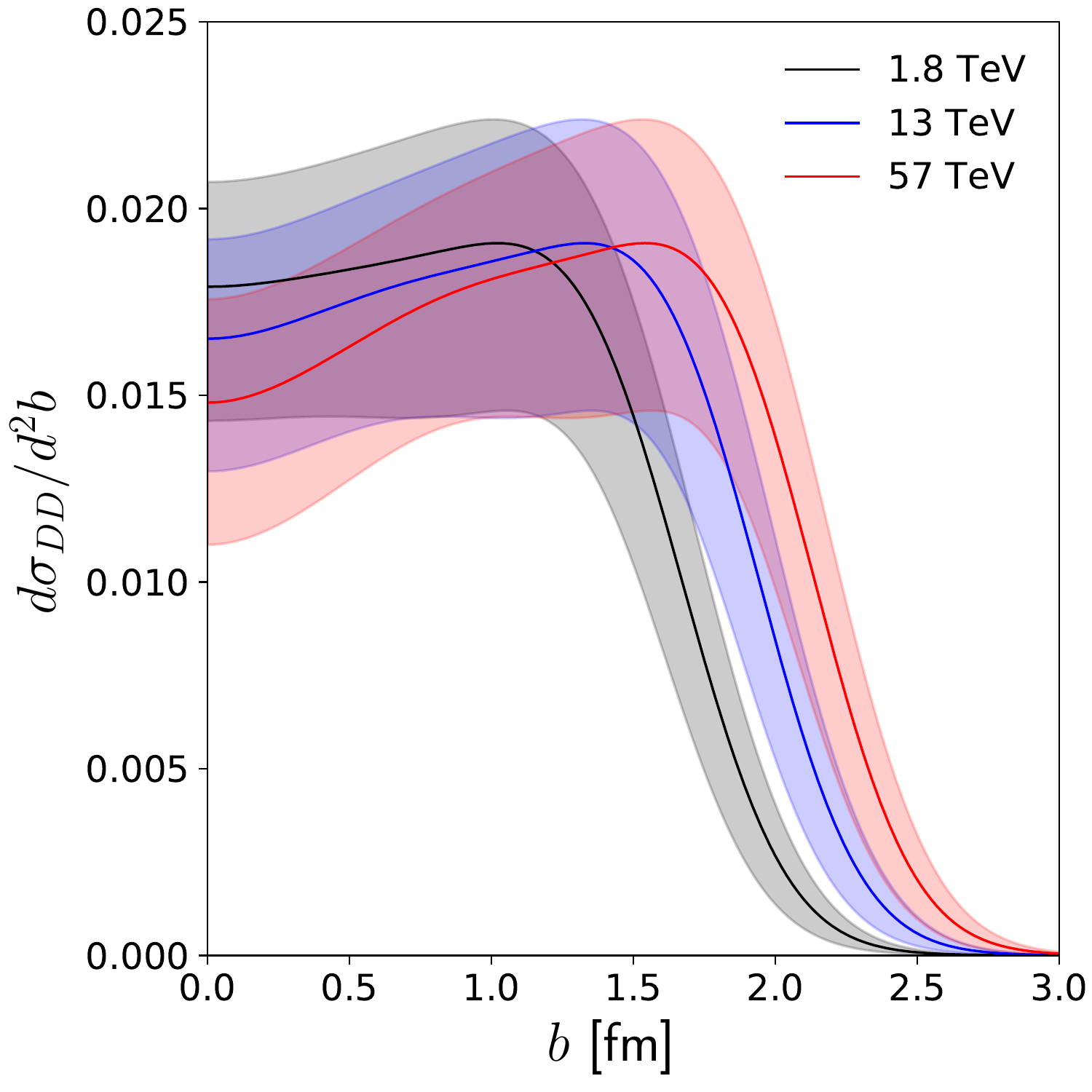}
  \caption{ Predictions for the energy dependence of the elastic (left panel), total single diffractive (center panel) and double diffractive (right panel) cross sections in impact parameter space. For each energy, the central curve corresponds to ${\cal{K}}=2.0$ and the band around it covers the region between the results for ${\cal{K}}=1.5$ and ${\cal{K}}=2.5$. 
}
  \label{fig:cross_sections_bspace}
\end{figure}

The predictions for  energy evolution of the cross sections in the impact parameter space are presented in  Fig.~\ref{fig:cross_sections_bspace}. 
The elastic, total single diffractive and double diffractive differential cross sections as a function of $b$ are presented  in the left, central and right panels of Fig.~\ref{fig:cross_sections_bspace}, respectively, from Tevatron to  cosmic rays energies. In contrast to the elastic scattering, which is mainly central and have a  magnitude that increases with the energy, approaching the black disk limit, the total single diffractive cross section becomes more peripheral with its maximum moving to larger impact parameter as the energy increases. In addition, the magnitude of the SD cross section at $b = 0$ decreases with the increasing of the center -- of -- mass energy. Similar behaviors are predicted by the updated version of the Miettinen -- Pumplin model presented in Ref.~\cite{nosMP}. The double diffractive cross section also becomes more peripheral with the increasing of energy, but the decreasing for central collisions is slower in comparison to the SD cross section.

\begin{figure}[t]
 \centering
   \includegraphics[scale=0.54]{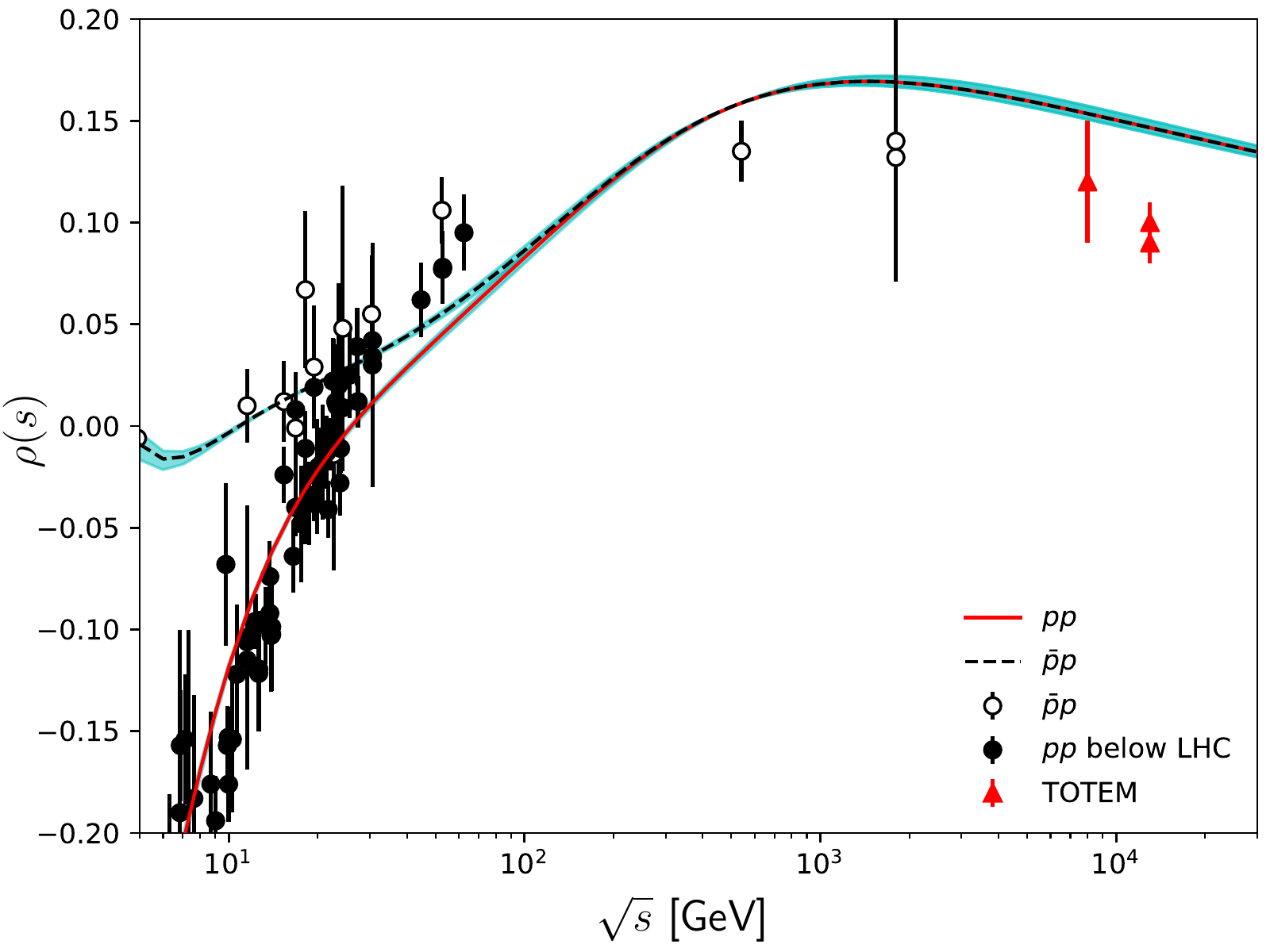}
    \includegraphics[scale=0.54]{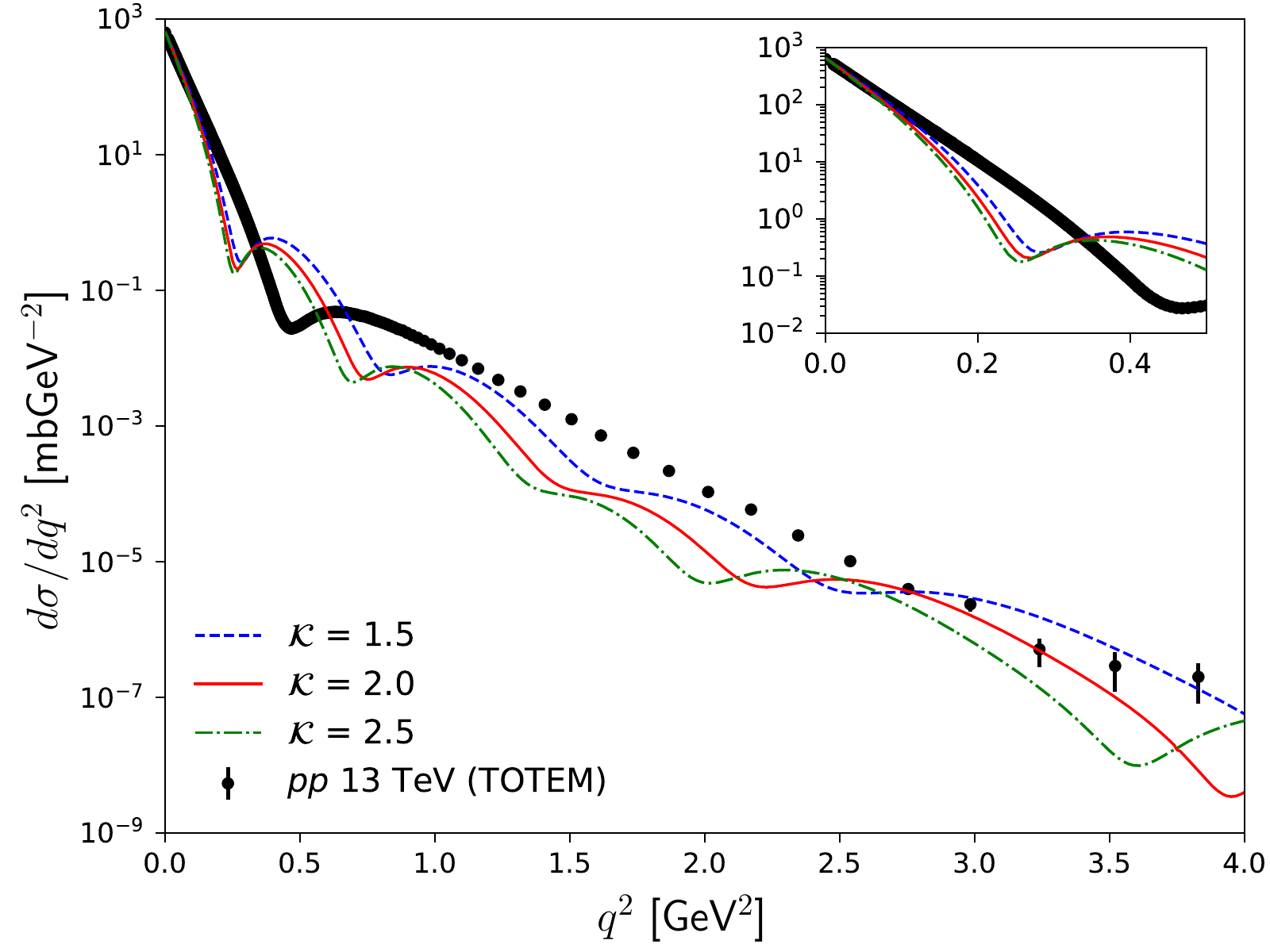}
  \caption{ Predictions for the $\rho$ parameter (left panel) and for the differential elastic cross section at 13 TeV (right panel). The central curve corresponds to ${\cal{K}}=2.0$ and the band around it covers the region between the results for ${\cal{K}}=1.5$ and ${\cal{K}}=2.5$. Data from \cite{pdg:2018,Antchev:2016,Antchev:2019a,Antchev:2019c}.}
  \label{fig:rho}
\end{figure}

During the last few years, the TOTEM Collaboration have released very precise data for the $\rho$ parameter and for the transverse momentum dependence of the elastic cross section. In particular, the experimental data for  $\rho$ have motivated an intense debate about the existence of a crossing-odd component, namely an Odderon. It is expected that such object may play an essential role in the description of high -- energy elastic scattering (See {\it e.g.}. Refs. 
  \cite{Martynov:2017zjz,Khoze:2018kna,Martynov:2018nyb,Troshin:2018ihb,Broniowski:2018xbg,Gotsman:2018buo,Goncalves:2018nsp,Martynov:2018sga,Csorgo:2018uyp}. In order to present our predictions for this quantity we need the real part of the elastic scattering amplitude. Therefore, we considered the single -- subtracted Integral Dispersion Relations (IDR)~\cite{Goldberg:1957,Soding:1964}.
\begin{align}
\frac{\Real F_+(s)}{s} & = \frac{{\cal{C}}}{s} + \frac{2s}{\pi} P \int_{s_\text{th}}^{\infty} ds' \left[\frac{1}{s'^2 - s^2}\right]\frac{\Imag F_+(s')}{s'},\label{eq:idr_even}\\
\frac{\Real F_-(s)}{s} & = \frac{2}{\pi} P \int_{s_\text{th}}^{\infty} ds' \left[\frac{s'}{s'^2 - s^2}\right]\frac{\Imag F_-(s')}{s'}\label{eq:idr_odd} \,\,,
\end{align}
\noindent where $s_\text{th} = 4m_p^2 = 3.521$~GeV$^2$ {represents the lower energy threshold}, ${\cal{C}}$ is the subtraction constant and $F_\pm$ are the  crossing even (+) and odd (-) amplitudes related to the physical amplitudes by
\begin{equation}
 F_{\pm} = \frac{1}{2}(F_{pp} \pm F_{\bar{p}p}). \label{eq:amp_evenodd}
\end{equation}
% In our calculations we will assume that ${\cal{C}} = 0$. 
% 
% NEW
% 
{ The value of the subtraction constant $\mathcal{C}$ is unknown. In principle, it can be determined in a global fit to $\sigma_{tot}$ and $\rho$ data, but that is not the aim of the present work. As the subtraction constant contributes only at low energies, we do not expect that it influences the  high energy predictions and extrapolations. Therefore, we will set $\mathcal{C}=0$ in what follows.
}
Our predictions for the $\rho$ parameter are presented in Fig.~\ref{fig:rho} (left panel). We { see} that the impact of the ${\cal{K}}$ factor in the result is small, with the predictions overestimating the TOTEM data at high energy. {Such result indicates that our assumptions for  the eikonal cross section  are very simplistic and should be improved taking into account of new contributions. Similar conclusion have been reached in recent studies performed using {\it e.g.} the DGM model \cite{Broilo:2019xhs}. One possibility is the inclusion of the Odderon contribution, which implies  distinct energy dependencies of the $pp$ and $p\bar{p}$ cross sections. In our model,  
the high energy behaviour of the predictions is determined by $\sigma_{pQCD}$. At high energies, this quantity is dominated by gluon -- gluon interactions, which implies that our model predicts identical behaviours for the $pp$ and $\bar{p}p$ cross sections. Therefore, in order to include the Odderon contribution in our model, the description of the soft part at high energies should be improved.} 
Our predictions for the transverse momentum dependence of the differential elastic cross section considering a $pp$ collision at $\sqrt{s} = 13$ TeV is presented in Fig.~\ref{fig:rho} (right panel). { Such quantity is sensitive to the description of the overlap function and, therefore, to the internal structure of the incident hadrons. Recent studies \cite{dre,tyu} have pointed out that in order to describe $d\sigma_{el}/dq^2$ at large and small values of the squared momentum transfer $q^2$, a multi-layer structure must be assumed for the proton, with the internal layers being accessed at large -- $q^2$. In our model we have assumed that the overlap function can be described in terms of the proton form factor, being the same for gluon -- gluon, gluon -- quark and quark -- quark interactions. Surely, such assumption is a very crude approximation for the spatial distribution of the quarks and gluons inside the proton. Therefore, it is natural to expect that our model will not be able to describe the experimental data for large momentum transfer. Such expectation is confirmed by the results presented in Fig.~\ref{fig:rho} (right panel), which demonstrate that  }the model is able to describe the experimental data for the differential elastic cross section at small values of the squared momentum transfer $q^2$,  but fails to describe the position of the dip and the behavior at large $q^2$. Similar results are obtained for other values of the center -- of -- mass energy. These two shortcomings point out possible directions for the improvement of the model discussed in this paper. In particular, the results for $d\sigma_{el}/dq^2$ indicate that the simplistic approach for the description of the spatial distribution of the partons inside the hadron, present in the Eq.~(\ref{eq:factb}), must be improved { {\it e.g.} assuming different overlap functions for the distinct subprocesses}. Such study is a work in progress. It is important to emphasize that similar conclusions were derived in the Refs.~\cite{lund,nosMP}, using distinct approaches to that used in this paper.

\section{Summary}
\label{sec:Sum}
 The description of the high energy regime of the hadronic interactions is one of the main challenges of the QCD. In particular, the treatment of the diffractive excitation is strongly dependent on the modelling of the internal degrees of freedom of the hadrons.  In this paper we have proposed a model to describe the eigenstates of interactions and the interaction between them, with the high energy behavior of the cross sections being determined by perturbative QCD. We have demonstrated that the model is able to describe the current data for the energy dependence of the elastic and total single diffractive cross sections and have presented our prediction for the double diffractive cross section, which can be tested by future experimental data. The impact parameter dependence of the cross sections was discussed. In addition, we have estimated the $\rho$ parameter and the differential elastic cross section and compared with the recent data. Such analysis demonstrated that although the model describe quite well the energy dependence of the cross sections, the description of the spatial distribution of the partons inside the hadron must be improved. A study about this topic is in progress.

%*****************************
\section*{Acknowledgements}
%*****************************
V. P. G. thank the members of the Faculty of Nuclear Sciences and Physical Engineering of the Czech Technical University in Prague and of the Nuclear Physics Institute ASCR by the warm hospitality during the completion of this work.
This work was  partially financed by the Brazilian funding
agencies CNPq,   FAPERGS and INCT-FNA (process number 
464898/2014-5 and 155628/2018-6).

%References

\end{document}